\documentclass[fleqn,usenatbib]{mnras}
\usepackage[T1]{fontenc}
\usepackage{ae,aecompl}

\usepackage{graphicx}	
\usepackage{amsmath}	
\usepackage{amssymb}	

\usepackage{ulem}

\usepackage{xcolor}
\usepackage{ragged2e}

\newcommand{\etal}{{et al.}}

\title[Core radio emissions from black hole sources] 
	{On the origin of core radio emissions from black hole sources in the realm of relativistic shocked accretion flow  }

\author[Das et al.]{Santabrata Das$^{1}$\thanks{E-mail: sbdas@iitg.ac.in (SD)} 
Anuj Nandi$^{2}$\thanks{E-mail: anuj@ursc.gov.in (AN)}, C. S. Stalin$^3$, Suvendu Rakshit$^{4}$,
\newauthor Indu Kalpa Dihingia$^{5}$, Swapnil Singh$^2$, Ramiz Aktar$^{6}$, Samik Mitra$^{1}$\\
$^1$ Indian Institute of Technology Guwahati, Guwahati, 781039, Assam, India\\
$^2$ Space Astronomy Group, ISITE Campus, U. R. Rao Satellite Center, Outer Ring Road, Marathahalli, Bangalore 560037, India\\
$^3$ Indian Institute of Astrophysics, Koramangala, Bangalore 560034, India\\
$^4$ Aryabhatta Research Institute of Observational Sciences, Manora Peak, Nainital 263002, India \\
$^5$ Discipline of Astronomy, Astrophysics and Space Engineering, Indian Institute of Technology Indore, Indore 453552, India\\
$^6$ Department of Astronomy, Xiamen University, Xiamen, Fujian 361005, People's Republic of China\\
}

\date{Accepted XXX. Received YYY; in original form ZZZ}

\begin{document}
\label{firstpage}
\pagerange{\pageref{firstpage}--\pageref{lastpage}}
\maketitle

\begin{abstract}
	
	We study the relativistic, inviscid, advective accretion flow around the black holes and investigate a key feature of the accretion flow, namely the shock waves. We observe that the shock-induced accretion solutions are prevalent and such solutions are commonly obtained for a wide range of the flow parameters, such as energy (${\cal E}$) and angular momentum ($\lambda$), around the black holes of spin value $0\le a_{\rm k} < 1$. When the shock is dissipative in nature, a part of the accretion energy is released through the upper and lower surfaces of the disc at the location of the shock transition. We find that the maximum accretion energies that can be extracted at the dissipative shock ($\Delta{\cal E}^{\rm max}$) are $\sim 1\%$ and $\sim 4.4\%$ for Schwarzschild black holes ($a_{\rm k}\rightarrow 0$) and Kerr black holes ($a_{\rm k}\rightarrow 1$), respectively. Using $\Delta{\cal E}^{\rm max}$, we compute the loss of kinetic power (equivalently shock luminosity, $L_{\rm shock}$) that is enabled to comply with the energy budget for generating jets/outflows from the jet base ($i.e.$, post-shock flow). We compare $L_{\rm shock}$ with the observed core radio luminosity ($L_R$) of black hole sources for a wide mass range spanning $10$ orders of magnitude with sub-Eddington accretion rate and perceive that the present formalism seems to be potentially viable to account $L_R$ of $16$ Galactic black hole X-ray binaries (BH-XRBs) and $2176$ active galactic nuclei (AGNs). We further aim to address the core radio luminosity of intermediate-mass black hole (IMBH) sources and indicate that the present model formalism perhaps adequate to explain core radio emission of IMBH sources in the sub-Eddington accretion limit.
	
\end{abstract}


\section{Introduction}
\label{sec:intro}

The observational evidence of the ejections of matter from the BH-XRBs \cite[]{Rodriguez-etal1992,Mirabel-Rodriguez1994} and AGNs \cite[]{Jennison-DasGupta1953,Junor-etal1999} strongly suggests that there possibly exists a viable coupling between the accreting and the outflowing matters \cite[]{Feroci-etal1999,Willott-etal1999,Ho-Peng2001,Pahari-etal2018,Russell-etal2019a,deHaas-etal2021}. Since the ejected matters are in general collimated, they are likely to be originated from the inner region of the accretion disc and therefore, they may reveal the underlying physical processes those are active surrounding the black holes. Further, observational studies indicate that there is a close nexus between the jet launching and the spectral states of the associated black holes \cite[]{Vadawale-etal2001,Chakrabarti-etal2002,Gallo-etal2003,Fender-etal2009,Radhika-etal2016,Blandford-etal2019}. All these findings suggest that the jet generation mechanism seems to be strongly connected with the accretion process around the black holes of different mass irrespective to be either BH-XRBs or AGNs. Meanwhile, numerous efforts were made both in theoretical  \cite[]{Chakrabarti1999,Das-Chakrabarti1999,Blandford-Begelman1999,Das-etal2001a,McKinney-Blandford2009,Das-etal2014,Ressler-etal2017,Aktar-etal2019,Okuda-etal2019} as well as observational fronts to explain the disc-jet symbiosis \cite[]{Feroci-etal1999,Brinkmann-etal2000,Nandi-etal2001,Fender-etal2004,Miller-Jones-etal2012,Miller-etal2012,Sbarrato-etal2014,Radhika-etal2016,Svoboda-etal2017,Blandford-etal2019}.

The first ever attempt to examine the correlation between the X-ray ($L_{\rm X}$) and radio ($L_{\rm R}$) luminosities for black hole candidate GX 339-4 during its hard states was carried out by \cite{Hannikainen-etal1998}, where it was found that $L_{\rm R}$ scales with $L_{\rm X}$ following a power-law. Soon after, \cite{Fender2001} reported that the compact radio emissions are associated with the Low/Hard State (LHS) of several black hole binaries. Similar trend was seen to follow by several such BH-XRBs \cite[]{Corbel-etal2003,Gallo-etal2003}. Later, \cite{Merloni-etal2003} revisited this correlation including the low-luminosity AGNs (LLAGNs) and found tight constraints on the correlation described as the {\it Fundamental Plane of the black hole activity} in a three-dimensional plane of ($L_{\rm R}, L_{\rm X}, M_{\rm BH}$), where $M_{\rm BH}$ denotes the mass of the black hole. Needless to mention that the above correlation study was conducted considering the core radio emissions at $5$ GHz in all mass scales ranging from stellar mass ($\sim 10 ~M_\odot$) to Supermassive ($\sim 10^{6-10} ~M_\odot$) black holes. To explain the correlation,  \cite{Heinz-Sunyaev2003} envisaged a non-linear dependence between the mass of the central black hole and the observed flux considering core dominated radio emissions. Subsequently, several group of authors further carried out the similar works to reveal the rigor of various physical processes responsible for such correlation \cite[]{Falcke-etal2004,Kording-etal2006,Merloni-etal2006,Wang-etal2006,Panessa-etal2007,Gultekin-etal2009,Plotkin-etal2012,Corbel-etal2013,Dong-Wu2015,Panessa-etal2015,Nisbet-Best2016,Gultekin-etal2019}. 

In the quest of the disc-jet symbiosis, many authors pointed out that the accretion-ejection  phenomenon is strongly coupled and advective accreting disc plays an important role in powering the jets/outflows \cite[and references therein]{Das-Chakrabarti1999,Blandford-Begelman1999,Chattopadhyay-etal2004,Aktar-etal2019}. In reality, an advective accretion flow around the black holes is necessarily transonic because of the fact that the infalling matter must satisfy the inner boundary conditions imposed by the event horizon. During accretion, rotating matter experiences centrifugal repulsion against gravity that yields a virtual barrier in the vicinity of the black hole. Eventually, such a barrier triggers the discontinuous transition of the flow variables to form shock waves \cite[]{Landau-Lifshitz1959,Frank-etal2002}. In reality, the downstream flow is compressed and heated up across the shock front that eventually generates additional entropy all the way up to the horizon. Hence, accretion solutions harboring shock waves are naturally preferred according to the 2nd law of thermodynamics \cite[]{Becker-Kazanas2001}. Previous studies corroborate the presence of hydrodynamic shocks \cite[][]{Fukue1987,Chakrabarti1989,Nobuta-Hanawa1994,Lu-etal1999,Fukumura-Tsuruta2004,Chakrabarti-Das2004,Moscibrodzka-etal2006,Das-Czerny2011,Aktar-etal2015,Dihingia-etal2019b}, and magnetohydrodynamic (MHD) shocks \cite[]{Koide-etal1998,Takahashi-etal2002,Das-Chakrabarti2007,Fukumura-Kazanas2007,Takahashi-Takahashi2010,Sarkar-Das2016,Fukumura-etal2016,Okuda-etal2019,Dihingia-etal2020} in both BH-XRB and AGN environments.

Extensive numerical simulations of the accretion disc independently confirm the formation of shocks as well \cite[]{Ryu-etal1997,Fragile-Blaes2008,Das-etal2014,Generozov-etal2014,Okuda-Das2015,Sukova-Janiuk2015,Okuda-etal2019,Palit-etal2020}. Due to the shock compression, the post-shock flow becomes hot and dense that results in a puffed up torus like structure which acts as the effective boundary layer of the black hole and is commonly called as post-shock corona (hereafter PSC). In general, PSC is hot enough ($T \gtrsim 10^{9}$ K) to deflect outflows which may be further accelerated by the radiative processes active in the disc \cite[]{Chattopadhyay-etal2004}. Hence, the outflows/jets are expected to carry a fraction of the available energy (equivalently core emission) at the PSC, which in general considered as the base of the outflows/jets \cite[]{Chakrabarti1999,Das-etal2001b,Chattopadhyay-Das2007,Das-Chattopadhyay2008,Singh-Chakrabarti2011,Sarkar-Das2016}. Becker and his collaborators showed that the energy extracted from the accretion flow via isothermal shock can be utilized to power the relativistic particles emanating from the disc \cite[]{Le-Becker2005,Becker-etal2008,Das-etal2009,Lee-Becker2020}. Moreover, magnetohydrodynamical study of the accretion flows around the black holes also accounts for possible role of shock as the source of high energy radiation  \cite[]{Nishikawa-etal2005,Takahashi-etal2006,Hardee-etal2007,Takahashi-Takahashi2010}.

An important generic feature of shock wave is that it is likely to be radiatively efficient. For that, shocks become dissipative in nature where an amount of accreting energy is escaped at the shock location through the disc surface resulting the overall reduction of downstream flow energy all the way down to the horizon. This energy loss is mainly regulated by a plausible mechanism known as the thermal Comptonization process \cite[and references therein]{Chakrabarti-Titarchuk1995,Das-etal2010}. Assuming the energy loss to be proportional to the difference of temperatures across the shock front, the amount of energy dissipation at the shock  can be estimated \cite[]{Das-etal2010}, which is same as the accessible energy at the PSC. A fraction of this energy could be utilized to produce and power outflows/jets as they are likely to originate from the PSC around the black holes \cite[]{Chakrabarti1999,Das-etal2001a,Aktar-etal2015,Okuda-etal2019}.

Being motivated with this appealing energy extraction mechanism, in this paper, we intend to study the stationary, axisymmetric, relativistic, advective accretion flow around the black holes in the realm of general relativity and self-consistently obtain the global accretion solutions containing dissipative shock waves. Such dissipative shock solution has not yet been explored in the literature for maximally rotating black holes having spin $a_{\rm k}\rightarrow 1$. We quantitatively estimate the amount of the energy released through the upper and lower surface of the disc at the shock location and show how the liberated energy affects the shock dynamics. We also compute the maximum available energy dissipated at the shock for $0 \le a_{\rm k} < 1$. Utilizing the usable energy available at the PSC, we estimate the loss of kinetic power (which is equivalent to shock luminosity) from the disc ($L_{\rm shock}$) which drives the jets/outflows. It may be noted that the kinetic power associated with the base of the  outflows/jets is interpreted as the core radio emission. Further, we investigate the observed correlation between radio luminosities and the black hole masses, spanning over ten orders of magnitude in mass for BH-XRBs as well as AGNs. We show that the radio luminosities in both {BH-XRBs and AGNs are in general much lower as compared to the possible energy loss at the PSC and therefore, we argue that the dissipative shocks seem to be potentially viable to account the energy budget associated with the core radio luminosities in all mass scales. Considering this, we aim to reveal the missing link between the BH-XRBs and AGNs in connection related to the jets/outflows.  Employing our model formalism, we estimate the core radio luminosity of the intermediate mass black hole (IMBH) sources in terms of the central mass.

The article is organized as follows: In Section \ref{sec:equ}, we describe our model and mention the governing equations. We present the solution methodology in Section \ref{sec:sol}. In Section \ref{sec:result}, we discuss our results in detail. In Section \ref{sec:implication}, we discuss the observational implications of our formalism to explain the core radio emissions from black holes in all mass scales. Finally, we present the conclusion in Section \ref{sec:con}.

\section{Assumptions and governing model equations}
\label{sec:equ}

We consider a steady, geometrically thin, axisymmetric, relativistic, advective accretion disc around a black hole. Throughout the study, we use a unit system as $G=M_{\rm BH}=c=1$, where $M_{\rm BH}$, $G$ and $c$ are the mass of the black hole, gravitational constant and speed of light, respectively. In this unit system, length and angular momentum are expressed in terms of $GM_{\rm BH}/c^2$ and $GM_{\rm BH}/c$. Since we have considered $M_{\rm BH} = 1$, the present analysis is applicable for black holes of all mass scales.

In this work, we investigate the accretion flow around a Kerr black hole and hence, we consider Kerr metric in Boyer-Lindquist coordinates \cite[]{Boyer-Lindquist1967} as,
$$\begin{aligned}
ds^2 = & ~g_{\mu\nu} dx^\mu dx^\nu,\\
= & ~g_{tt}dt^2 + 2g_{t\phi}dtd\phi + g_{rr} dr^2 + g_{\theta\theta} 
d\theta^2 + g_{\phi\phi} d\phi^2,
\end{aligned}\eqno(1)$$
where $x^\mu  ~(\equiv t,r,\theta,\phi)$ denote coordinates and
$g_{tt} = -(1 - 2r/\Sigma)$, $g_{t\phi} = -2 a_{\rm k}r\sin^2\theta/\Sigma$, 
$g_{rr}=\Sigma/\Delta$, $g_{\theta\theta}=\Sigma$ and $g_{\phi\phi}=A\sin^2\theta/\Sigma$ 
are the non-zero metric components. Here, $A=(r^2 + a_{\rm k}^2)^2 - \Delta a_{\rm k}^2
\sin^2\theta$, $\Sigma = r^2 + a_{\rm k}^2\cos^2\theta$, $\Delta = r^2 - 2r + a_{\rm k}^2$, and $a_{\rm k}$ is the black hole spin.
In this work, we follow a convention where the four velocities satisfy $u_\mu u^\mu=-1$.

Following \cite{Dihingia-etal2019a}, we obtain the governing equations that describe the accretion flow for a geometrically thin accretion disc which are given by,

(a) the radial momentum equation:
$$
\begin{aligned}
&u^ru^r_{,r} + \frac{1}{2}g^{rr}\frac{g_{tt,r}}{g_{tt}} + \frac{1}{2}u^ru^r
\left(\frac{g_{tt,r}}{g_{tt}} + g^{rr}g_{rr,r}\right)\\
+ &u^\phi u^tg^{rr}\left(\frac{g_{t\phi}}{g_{tt}}
g_{tt,r} - g_{t\phi,r}\right)
+\frac{1}{2}u^{\phi}u^{\phi}g^{rr}\left(\frac{g_{\phi\phi}g_{tt,r}}{g_{tt}} - g_{\phi\phi,r}\right)\\
&+ \frac{(g^{rr} + u^ru^r)}{e+p}p_{,r}=0.\\
\end{aligned}
\eqno(2)
$$

\noindent (b) the continuity equation:
$$
\dot{M} = -4\pi r u^r \rho H,
\eqno(3)
$$
where $e$ is the energy density, $p$ is the local gas pressure, $\dot{M}$ is the accretion rate treated as global constant, and $r$ stands for radial coordinate. Moreover, $H$ refers the local half-thickness of the disc and is given by \cite[]{Riffert-Herold1995,Peitz-Appl1997,Dihingia-etal2019a},
$$
H = \left(\frac{pr^3}{\rho \mathcal{F}} \right)^{1/2}; \qquad {\rm with}~ 
\mathcal{F}=\gamma_\phi^2\frac{(r^2 + a_{\rm k}^2)^2 + 2\Delta a_{\rm k}^2}
{(r^2 + a_{\rm k}^2)^2 - 2\Delta a_{\rm k}^2},
$$
where $\gamma_\phi^2=1/(1-v_\phi^2)$ is the bulk azimuthal Lorentz factor and $v_\phi^2 = u^\phi u_\phi/(-u^t u_t)$. We define the radial three velocity in the co-rotating frame as $v^2 = \gamma_\phi^2v_r^2$ and thus, we have the bulk radial Lorentz factor $\gamma^2_v = 1/(1-v^2)$, where $v_r^2 = u^ru_r/(-u^tu_t)$.

In order to solve equations (2-3), a closure equation in the form of Equation of State (EoS) describing the relation among the thermodynamical quantities, namely density ($\rho$), pressure ($p$) and energy density ($e$) is needed. For that we adopt an EoS for relativistic fluid which is given by \cite[]{Chattopadhyay-Ryu2009},
$$
e = \frac{\rho f}{\left(2 -\frac{m_p}{m_e}\right)},
$$
with
$$
f=\left[ 1+ \Theta \left( \frac{9\Theta +3}{3 \Theta +2}\right) \right] + \left[ \frac{m_p}{m_e} + \Theta \left( \frac{9\Theta m_e +3m_p}{3 \Theta m_e+ 2 m_p}\right) \right],
$$
where $\Theta~(=k_{\rm B}T/m_e c^2)$ is the dimensionless temperature, $m_e$ is the mass of electron, and $m_p$ is the mass of ion, respectively. According to the relativistic EoS, we express the speed of sound as $a_s = \sqrt{2\Gamma \Theta/(f+2\Theta)}$, where $\Gamma = (1+N)/N$ is the adiabatic index, and $N=(1/2)(df/d\Theta)$ is the polytropic index of the flow \cite[]{Dihingia-etal2019a}.

In this work, we use a stationary metric $g^{\mu\nu}$ which has axial symmetry and this enables us to construct two Killing vector fields $\partial_t$ and $\partial_\phi$ that provide two conserved quantities for the fluid motion in this gravitational field and are given by, 
$$
hu_\phi = {\rm constant}; \qquad -hu_t={\rm constant} = {\cal E},
\eqno(4)
$$
where $h~[=(e+p)/\rho]$ is the specific enthalpy of the fluid, ${\cal E}$ is the relativistic Bernoulli constant ($i. e.$, the specific energy of the flow). Here,  
$u_t = -\gamma_v \gamma_\phi/\sqrt{\lambda g^{t\phi} - g^{tt}}$, where
$\lambda ~(= -u_\phi/u_t$) denotes the conserved specific angular momentum.

\section{Solution Methodology}

\label{sec:sol}

We simplify equations (2) and (3) to obtain the wind equation in the co-rotating frame as,
$$
\frac{dv}{dr}= \frac{\mathcal{N}}{\mathcal{D}},
\eqno(5)
$$
where the numerator ${\cal N}$ is given by,
$$\begin{aligned}
\mathcal{N}=& - \frac{1}{r(r-2) } + \gamma_\phi^2 \lambda \frac{2a_{\rm k}}{r^2\Delta} 
+ \gamma_\phi^2\frac{4a_{\rm k}^2}{r^2\Delta(r-2)}\\
& -\gamma_\phi^2\Omega\lambda\frac{ 2a_{\rm k}^2- r^2(r-3)}{r^2\Delta}
+ 2a_{\rm k}\gamma_\phi^2\Omega\frac{ r^2(r-3)-2 a_{\rm k}^2}{r^2 \Delta(r-2)}\\
&+ \frac{2a_s^2}{\Gamma + 1}\bigg[ \frac{\left(r-a_{\rm k}^2\right)}{r\Delta} 
+ \frac{5}{2r} - \frac{1}{2\mathcal{F}}\frac{d\mathcal{F}}{dr}\bigg],\\
\end{aligned}\eqno(6)$$
and the denominator ${\cal D}$ is given by,
$$
\mathcal{D} = \gamma_v^2\bigg[v- \frac{2a_s^2}{v(\Gamma +1)}\bigg],
\eqno(7)
$$
where, $\Omega=u^\phi/u^t$ is the angular velocity of the flow.

Following \cite{Dihingia-etal2019a}, we obtain the temperature gradient as,
$$
\frac{d\Theta}{dr}=-\frac{2\Theta}{2N + 1}\bigg[\frac{\left(r-a_{\rm k}^2\right)}{r\Delta}
+\frac{\gamma_v^2}{v}\frac{dv}{dr} + \frac{5}{2r} - 
\frac{1}{2\mathcal{F}}\frac{d\mathcal{F}}{dr}\bigg].
\eqno(8)
$$

In order to obtain the accretion solution around the black hole, we solve equations (5-8) following the methodology described in \cite{Dihingia-etal2019a}. While doing this, we specifically confine ourselves to those accretion solutions that harbor standing shocks \cite[][]{Fukue1987,Chakrabarti1989,Yang-Kafatos1995,Lu-etal1999,Chakrabarti-Das2004,Fukumura-Tsuruta2004,Das2007,Chattopadhyay-Kumar2016,Sarkar-Das2016,Dihingia-etal2019b}. In general, during the course of accretion, the rotating infalling matter experiences centrifugal barrier at the vicinity of the black hole. Because of this, matter slows down and piles up causing the accumulation of matter around the black hole. This process continues until the local density of matter attains its critical value and once it is crossed, centrifugal barrier triggers the transition of the flow variables in the form of shock waves. In reality, shock induced global accretion solutions are potentially favored over the shock free solutions as the entropy content of the former type solution is always higher \cite[]{Das-etal2001a,Becker-Kazanas2001}. At the shock, the kinetic energy of the supersonic pre-shock flow is converted into thermal energy and hence, post-shock flow becomes hot and behaves like a Compton corona \cite[]{Chakrabarti-Titarchuk1995,Iyer-etal2015,Nandi-etal2018,Aktar-etal2019}. As there exists a temperature gradient across the shock front, it enables a fraction of the available thermal energy to dissipate away through the disc surface. Evidently, the energy accessible at the 
post-shock flow is same as the available energy dissipated at the shock. A part of this energy is utilized in the form of high energy radiations, namely the gamma ray and the X-ray emissions, and the rest is used for the jet/outflow generation as they are expected to be launched from the post-shock region \cite[]{Chakrabarti1999,Becker-etal2008,Das-etal2009,Becker-etal2011,Sarkar-Das2016}. These jets/outflows further consume some energy simultaneously for their thermodynamical expansion and for the work done against gravity. The remaining energy is then utilized to power the jets/outflows.

It may be noted that for radiatively inefficient adiabatic accretion flow, the specific energy in the pre-shock as well as post-shock flows remains conserved. In reality, the energy flux across the shock front becomes uniform when the shock width is considered to be very thin and the shock is non-dissipative \cite[]{Chakrabarti1989,Frank-etal2002} in nature.
However, in this study, we focus on the dissipative shocks where a part of the accreting energy is released vertically at the shock causing a reduction of specific energy in the post-shock flow. The mechanism by which the accreting energy could be dissipated at the shock is primarily governed by the thermal Comptonization process \cite[][]{Chakrabarti-Titarchuk1995} and because of this, the temperature in the post-shock region is decreased. Considering the above scenario, we model the loss of energy ($\Delta {\cal E}$) to be proportional to the temperature difference across the shock front and $\Delta {\cal E}$ is estimated as \cite[and references therein]{Das-etal2010,Singh-Chakrabarti2011,Sarkar-Das2016},
$$
\Delta {\cal E} = \beta (N_{+}a_{s+}^2 - N_{-}a_{s-}^2),
\eqno(9)
$$
where $\beta$ is the proportionality constant that accounts the fraction of the accessible thermal energy across the shock front. 
Here, the quantities expressed using the subscripts `$-$' and `$+$' refer their immediate pre-shock and post-shock values, respectively. Needless to mention that because of the energy dissipation at the shock, the post-shock flow energy (${\cal E}_{+}$) can be expressed as ${\cal E}_{+}= {\cal E}_{-} - \Delta {\cal E}$, where ${\cal E}_{-}$ denotes the energy of the pre-shock flow. In this work, we treat ${\cal E}_{-}$ and ${\cal E}_{+}$ as free parameters and applying them, we calculate $\Delta {\cal E}$ from the shocked accretion solutions. Needless to mention that the post-shock flow may become bound due to the energy dissipation ($\Delta {\cal E} > 0$) across the shock front, however, all the solutions under consideration are chosen as unbound in the pre-shock domain. With this, we calculate $\beta$ using equation (9) for shocked accretion solutions that lies in the range $0 < \beta < 1$.

It is noteworthy that in this work, the global accretion solutions containing shocks are independent of the accretion rate as radiative cooling processes are not taken into account for simplicity. This eventually imposes limitations in explaining the physical states of the accretion flow although the model solutions are suffice to characterize the accretion flow kinematics in terms of the conserved quantities, namely energy and angular momentum of the flow.
	
Now, based on the above insight on the energy budget, the total usable energy available in the post-shock flow is $\Delta {\cal E}$. Keeping this in mind, we calculate the loss of kinetic power by the disc corresponding to $\Delta {\cal E}$ in terms of the observable quantities and obtain the shock luminosity \cite[]{Le-Becker2004,Le-Becker2005} as,
$$
L_{\rm shock}=\dot M \times \Delta {\cal E} \times c^2~~{\rm erg~s^{-1}},
\eqno(10)
$$
where $L_{\rm shock}$ is the shock luminosity and $\dot M$ is the accretion rate. With this, we compute $L_{\rm shock}$ considering the dissipative shock mechanism and compare it with core radio luminosity observed from the black hole sources. Indeed, it is clear from equation (10) that $L_{\rm shock}$ may be degenerate due to the different combinations of $\dot M$ and $\Delta {\cal E}$. In this work, we choose the spin value of the black hole in the range $0 \le a_{\rm k} \le 0.99$. Moreover, in order to represent the LHS of the black hole sources (as `compact' jets are commonly observed in the LHS \cite[]{Fender-etal2004}), we consider the value of accretion rate in the range ${\dot m} ={\dot M}/\dot {M}_{Edd}= 10^{-5} - 1.0$ \cite[]{Wu-Liu2004,Athulya-etal2021}, where $\dot M_{Edd}$ is the Eddington mass accretion rate and is given by ${\dot M}_{Edd} = 1.39 \times 10^{17} \left(M_{\rm BH}/M_\odot\right)~{\rm g~sec^{-1}}$. Furthermore, in order to examine the robustness of our model formalism, we vary the mass of the central black hole in a wide range starting form stellar mass to Supermassive scale, and finally compare the results with observations.

\section{Results}

\label{sec:result}

\begin{figure}
	\begin{center}
		\includegraphics[scale=0.425]{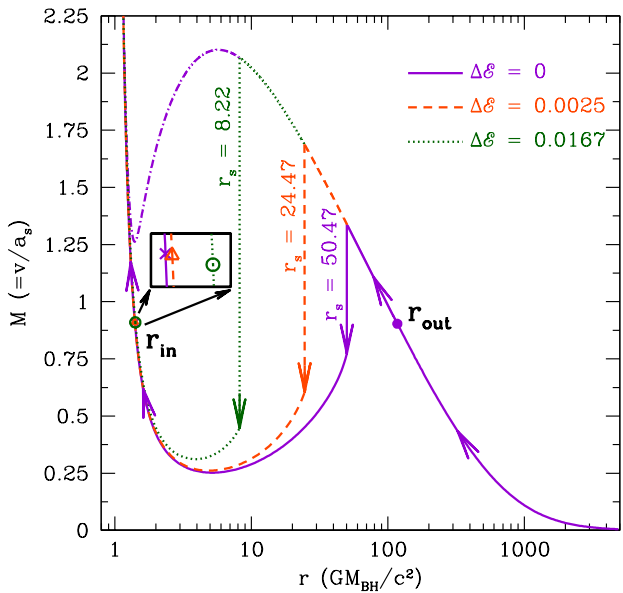}
	\end{center}
		\caption{Plot of Mach number ($M=v/a_s$) as function of radial coordinate (r). Here, the flow parameters are chosen as ${\cal E_{-}} = 1.002$ and $\lambda = 2.01$, and black hole spin is considered as $a_{\rm k}=0.99$. Results depicted with solid (purple), dashed (orange) and dotted (green) curves are obtained for $\Delta {\cal E} = 0, 0.0025$, and $0.0167$, respectively. At the inset, inner critical points ($r_{\rm in}$) are zoomed which are shown using open circle, open triangle and a cross whereas outer critical point ($r_{\rm out}$) is shown using filled circle. Vertical arrows represent the locations of the shock transition ($r_{\rm s}$) and the arrows indicate the overall direction of flow motion towards the black hole. See text for details. 
		}
	\label{fig:r_M_dE}
\end{figure}

In Fig. \ref{fig:r_M_dE}, we depict the typical accretion solutions around a rotating black hole of spin $a_{\rm k} = 0.99$. In the figure, we plot the variation of Mach number ($M=v/a$) as function of radial coordinate ($r$). Here, the flow starts its journey from the outer edge of the disc at $r_{\rm edge}= 5000$ subsonically with energy ${\cal E}_{-}=1.002$ and angular momentum $\lambda= 2.01$. As the flow moves inward, it gains radial velocity due to the influence of black hole gravity and smoothly makes sonic state transition while crossing the outer critical point at $r_{\rm out}=117.5285$. At the supersonic regime, rotating flow experiences centrifugal barrier against gravity that causes the accumulation of matter in the vicinity of the black hole. Because of this, matter locally piles up resulting the increase of density. Undoubtedly, this process is not continued indefinitely due to the fact that at the critical limit of density, centrifugal barrier triggers the discontinuous transition in the flow variables in the form of shock waves \cite[]{Fukue1987,Frank-etal2002}. At the shock, supersonic flow jumps into the subsonic branch where all the pre-shock kinetic energy of the flow is converted into thermal energy. In this case, the flow experiences shock transition at $r_s=50.47$. Just after the shock transition, post-shock flow momentarily slows down, however gradually picks up its velocity and ultimately enters into the black hole supersonically after crossing the inner critical point smoothly at $r_{\rm in}=1.4031$. This global shocked accretion solution is plotted using solid (purple) curve where arrows indicate the direction of the flow motion and the vertical arrow indicates the location of the shock transition. Next, when a part of the flow energy ($\Delta {\cal E}$) is radiated away through the disc surface at the shock, the post-shock thermal pressure is reduced and the shock front is being pushed further towards the horizon. Evidently, the shock settles down at a smaller radius in order to maintain the pressure balance across the shock front. Following this, when $\Delta {\cal E}= 0.0025$ is chosen, we obtain $r_s = 24.47$ and $r_{\rm in}= 1.4047$, and the corresponding solution is plotted using the dashed curve (orange). When the energy dissipation is monotonically increased, for the same set of flow parameters, we find the closest standing shock location at $r_s = 8.22$ for $\Delta {\cal E}= 0.0167$. This solution is presented using dotted curve (green) where $r_{\rm in}= 1.4147$. For the purpose of clarity, in the inset, we zoom the inner critical point locations as they are closely separated. In the figure, critical points and the energy dissipation parameters are marked. What is more is that following \cite{Chakrabarti-Molteni1993,Yang-Kafatos1995,Lu-etal1997,Fukumura-Kazanas2007}, the stability of the standing shock is examined, where we vary the shock front radially by an infinitesimally small amount in order to perturb the radial momentum flux density ($T^{rr}$, \cite{Dihingia-etal2019a}). When shock is dynamically stable, it must come back to its original position and the criteria for stable shock is given by, $\kappa(r_{s}) = \left( \frac{dT^{rr}_2}{dr}-\frac{dT^{rr}_1}{dr}\right) < 0$ \cite[]{Fukumura-Kazanas2007}. Invoking this criteria, we ascertain that all the standing shocks presented in Fig. \ref{fig:r_M_dE} are stable. For the same shocked accretion solutions, we compute the various shock properties \cite[see][]{Das2007,Das-etal2009}, namely, shock location ($r_s$), compression ratio ($R$), shock strength ($S$), scale height ratio ($H_{+}/H_{-}$), and present them in Table \ref{tab:sok}. In reality, as $\Delta{\cal E}$ increases, shock settles down at the lower radii (Fig. \ref{fig:r_M_dE}) and hence, the temperature of PSC increases due to enhanced shock compression. Moreover, since the disk thickness is largely depends on the local temperature, the scale height ratio increases with the increase of $\Delta{\cal E}$ yielding the PSC to be more puffed up for stronger shock. Accordingly, we infer that geometrically thick PSC seems to render higher energy dissipation (equivalently $L_{\rm shock}$) that possibly leads to produce higher core radio luminosity.
	
\begin{table}
	\caption{Various shock properties computed for solutions presented in Fig. \ref{fig:r_M_dE}, where $\lambda = 2.01$, ${\cal E}_{-}=1.002$ are chosen. See the text for details.
	}
	\centering
	\begin{tabular}{@{}lcccccc} \hline \hline
	$\Delta{\cal E}$ &  $r_{\rm out}$ & $r_{\rm in}$ & $r_s$ & $R$ & $S$ & $H_{+}/H_{-}$ \\ \hline
		 0.0    & 117.5285 & 1.4031 & 50.47 & 1.58 & 1.75 & 1.11 \\
		 0.0025 & 117.5285 & 1.4047 & 24.47 & 2.43 & 2.86 & 1.20 \\
		 0.0167 & 117.5285 & 1.4147 &  8.22 & 3.67 & 4.35 & 1.26 \\ \hline
	\end{tabular}
	\justify
	\noindent $Note$: $\Delta{\cal E}$ is the energy loss, $r_{\rm out}$ is the outer critical point, $r_{\rm in}$ is the inner critical point, $r_s$ is the shock location, $R$ is the compression ratio, $S$ is the shock strength, and $H_{+}/H_{-}$ refers scale height ratio.
	
	\label{tab:sok}
\end{table}

\begin{figure}
	\begin{center}
		\includegraphics[scale=0.4]{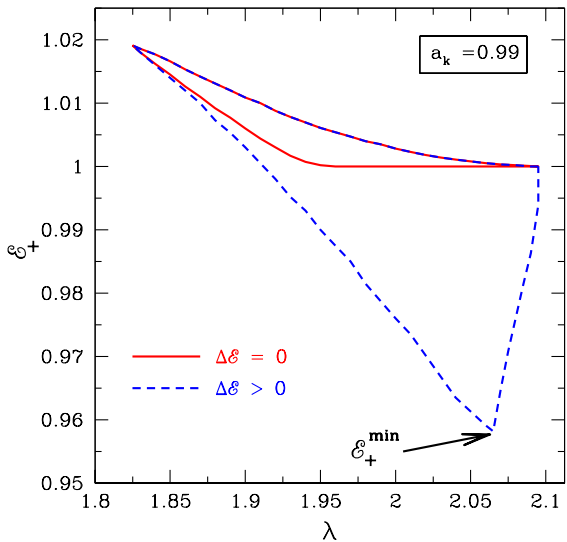}
	\end{center}
	\caption{Plot of parameter space in $\lambda-{\cal E}_{+}$ plane that admitted shock induced global accretion solutions around the black holes. For fixed $a_{\rm k}=0.99$, we obtain shocked accretion solution passing through the inner critical point and having the minimum energy ${\cal E}^{\rm min}_{+}$. The maximum amount of energy is lost by the flow via the disc surface at the shock for ${\cal E}^{\rm min}_{+}$. See text for details. 
		}
	\label{fig:amda_Ein_Eout}
\end{figure}

We examine the entire range of ${\cal E}_{+}$ and $\lambda$ that provides the global transonic shocked accretion solution around a rapidly rotating black hole of spin value $a_{\rm k}=0.99$. The obtained results are presented in Fig. \ref{fig:amda_Ein_Eout}, where the effective region bounded by the solid curve (in red) in $\lambda-{\cal E}_{+}$ plane provides the shock solutions for $\Delta{\cal E}=0$. Since energy dissipation at shock is restricted, the energy across the shock front remains same that yields ${\cal E}_{-} = {\cal E}_{+}$. When energy dissipation at shock is allowed ($i.e., \Delta{\cal E}> 0$), we have ${\cal E}_{+} < {\cal E}_{-}$ irrespective to the choice of $\lambda$ values. We examine all possible range of $\Delta{\cal E}$ that admits shock solution and separate the domain of the parameter space in $\lambda-{\cal E}_{+}$ plane using dashed curve (in blue). Further, we vary $\lambda$ freely and calculate the minimum flow energy with which flow enters into the black hole after the shock transition. In absence of any energy dissipation between the shock radius ($r_{\rm s}$) and horizon ($r_{\rm h}$), $i.e.,$ in the range $r_h < r <r_s$, this minimum energy is identical to the minimum energy of the post-shock flow (${\cal E}_{+}$) and we denote it as ${\cal E}^{\rm min}_{+}$. Needless to mention that ${\cal E}^{\rm min}_{+}$ strongly depends on the spin of the black hole ($a_{\rm k}$) marked in the figure. It is obvious that for a given $a_{\rm k}$, the maximum energy that can be dissipated at the shock is calculated as $\Delta{\cal E}^{\rm max} = {\cal E}_{-} - {\cal E}^{\rm min}_{+}$.  

\begin{figure}
	\begin{center}
		\includegraphics[scale=0.4]{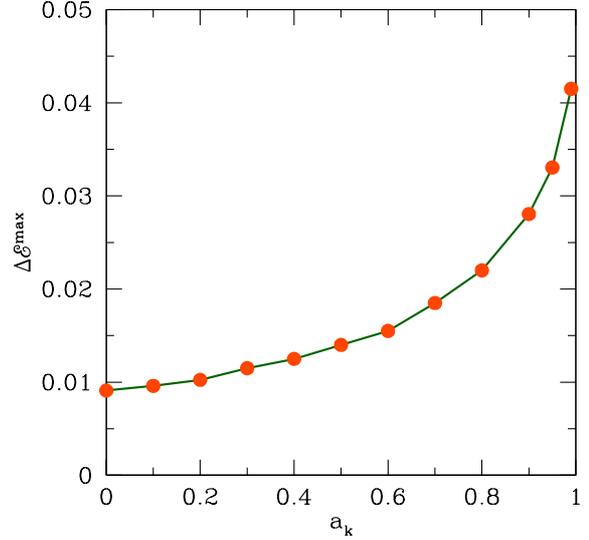}
	\end{center}
	\caption{Plot of maximum available energy across the 
		shock front ($\Delta {\cal E}^{\rm max}$) as function of the black hole spin ($a_{\rm k}$). Obtained results are depicted by the filled circles in orange color which are joined by the green lines. See text for details. 
		}
		\label{fig:ak_dEmax}
\end{figure}

Subsequently, we freely vary all the input flow parameters, namely ${\cal E}_{-}$ and $\lambda$, and calculate $\Delta {\cal E}^{\rm max}$ {for a given $a_{\rm k}$. The obtained results are presented in Fig. \ref{fig:ak_dEmax}, where we depict the variation of $\Delta {\cal E}^{\rm max}$ as function $a_{\rm k}$. We find that around $1\%$ of the flow energy can be extracted at the dissipative shock for Schwarszchild black hole (weakly rotating, $a_{\rm k} \rightarrow 0$) and about $4.4\%$ of the flow energy can be extracted for Kerr black hole (maximally rotating, $a_{\rm k} \rightarrow 1$). 

\begin{table*}
	\caption{Physical and observable parameters of BH-XRBs. Core radio luminosities ($L_R$) are complied from the literature for several sources, if available. For the rest, $L_R$ is calculated using source distance ($D$), observation frequency ($\nu$) and core radio flux ($F_5$) values using $L_R = 4 \pi \nu F_5 D^2$, where $D$ refers source distance.}
	\centering{
	\begin{tabular}{@{}lcccccccccc} \hline \hline
		Source Name       & Mass & Distance & Spin  & $\nu$ & Radio Flux & Core radio luminosity & References\\ 
		      &   ($M_{\rm BH}$) & ($D$) & ($a_{\rm k}$) & & ($F_{5}$) & at $5$ GHz ($L_{\rm R}$)& \\
		                 & (in $M_\odot$)  & (in kpc) & & (in GHz) & (in mJy) & (in $10^{30}$ erg s$^{-1}$)& \\ \hline
	    4U 1543-47        & $9.42\pm0.97$  & $7.5\pm1.0$    & $\sim 0.85$ & $4.8$ & $3.18-4.00$ & $1.03-1.29$ & $1,2,3,44$ \\ \hline
		Cyg X-1           & $14.8\pm1.0$   & $1.86\pm0.12$  & $>0.99$ & $15$  & $6.00-19.60$ & $0.124-0.406$ & $4,5,3,45ab$\\ \hline
		GRO J1655-40      & $6.3\pm0.25$   &  $3.2\pm0.2$   & $\sim0.98$ & $4.86$ & $1.46-2.01$ & $0.087-0.120$ & $6,7,3,46$\\ \hline
        GRS 1915+105      & $12.4^{+2.0}_{-1.8}$ & $8.6^{+2.0}_{-1.6}$ & $\sim0.99$ & $5.0$ &$25.75-198.77$ & $11.396-87.967$  & $8,8,3,47$\\ \hline
        XTE J1118+480      & $7.1\pm1.3$    & $1.8\pm0.6$   & --- & $15$  & $6.2-7.5$  & $0.069-0.084$ & $9,9,3$\\ \hline 
        XTE J1550-564     & $9.1\pm0.6$     & $4.4\pm0.5$    & $\sim0.78$ & $4.8$ & $0.88-7.45$  & $0.098-0.829$ & $10,10,3,48$\\ \hline
		Cyg X-3 & $2.4^{+2.1}_{-1.1}~^\dagger$ & $7.4\pm1.1$    & --- &  ---     & ---       & $4.36-269.15$ & $11,12,13$\\ \hline
        GX 339-4          & $10.08^{+1.81}_{-1.80}$    & $8.4\pm0.9$   & $>0.97$ &   ---    &  ---      & $0.00178-0.8128$ & $14,15,13,49$\\ \hline
        XTE J1859+226     & $6.55\pm1.35$       & $6-11$   & $\sim0.6$ &  ---    &   ---     & $0.151-0.199$ & $16,17ab,13,50$\\ \hline
        H 1743-322        & $11.21^{+1.65}_{-1.96}$         & $8.5\pm0.8$   & $<0.7$ & $4.8$ & $0.12-2.37$ & $0.05-0.984$ & $18,19,20,19$\\ \hline
		IGR J17091-3624   & $10.6-12.3$     & $11-17$   & $<0.27$ &  $5.5$    & $0.17-2.41$ & $0.18-2.52$ & $21,22,22,51$ \\ \hline
        4U 1630-472       & $10.0\pm0.1$    & $11.5\pm0.3$   & $\sim0.98$ & $4.86^{~\boxdot}$  & $1.4\pm0.3$  & $1.08-1.98$  & $23, 24, 25,52ab$\\
		                  &       &          & &$4.80^{~\boxtimes}$ & $2.6\pm0.3$  & & \\ \hline
		MAXI J1535-571    & $6.47^{+1.36}_{-1.33}$ & $4.1^{+0.6}_{-0.5}$    & $\sim0.99$ & $5.5$ & $0.18-377.20$ & $0.02-41.74$    & $26,27,28,53$\\ \hline
        MAXI J1348-630    & $11\pm2$   & $2.2^{+0.5}_{-0.6}$    & --- & $5.5$ & $3.4\pm0.2$& $0.108$ & $29,30,31$\\ \hline
        MAXI J1820+070    & $5.73-8.34$ & $2.96\pm0.33$    & $\sim 0.2$ & $4.7$ & $62\pm4$   & $3.06$ & $32, 33, 34,54$\\ \hline
        V404 Cyg & $9.0^{+0.2}_{-0.6}$ & $2.39 \pm 0.14$ & $> 0.92$  & $4.98^{~\boxplus}$ & $0.141-0.680$ &$0.005-0.023$& $35,36, 37, 55$ \\ \hline
%
        Swift J1357.2-0933 & $>9.3$          & $2.3-6.3$ & --- &   $5.5$    &   ---     & $0.0043-0.033$ & $38,39ab,40$\\ \hline
        MAXI J0637-430    & $8.0^\dagger$           & $10.0$   & --- & $5.5$ & $0.066\pm0.015$ & $0.043$ & $41,42,43$\\
        \hline                  

	\end{tabular}
} 
\\
    \justify
	\noindent{\bf References:} $1$: \cite{Orosz-2003}, $2$: \cite{Park-etal2004}, $3$: \cite{Gultekin-etal2019}, $4$: \cite{Orosz-etal2011a}, $5$: \cite{Reid-etal2011}, $6$: \cite{Greene-etal2001}, $7$: \cite{Jonker-Nelemans2004}, $8$: \cite{Reid-etal2014}, $9$: \cite{McClintock-etal2001}, $10$: \cite{Orosz-etal2011b}, $11$: \cite{Zdziarski-etal2013}, $12$: \cite{McCollough-etal2016}, $13$: \cite{Merloni-etal2003}, $14$: \cite{Sreehari-etal2019b}, $15$: \cite{Parker-etal2016}, $16$: \cite{Nandi-etal2018}, $17a$: \cite{Hynes-etal2002}, $17b$: \cite{Zurita-etal2002}, $18$: \cite{Molla-etal2017}, $19$: \cite{Steiner-etal2012}, $20$: \cite{Corbel-etal2005}, $21$: \cite{Iyer-etal2015}, $22$: \cite{Rodriguez-etal2011}, $23$: \cite{Seifina-etal2014}, $24$: \cite{Kalemci-etal2018}, $25$: \cite{Hjellming-etal1999}, $26$: \cite{Sreehari-etal2019}, $27$: \cite{Chauhan-etal2019}, $28$: \cite{Russell-etal2019a}, $29$: \cite{Lamer-etal2020}, $30$: \cite{Chauhan-etal2021}, $31$: \cite{Russell-etal2019b}, $32$: \cite{Torres-etal2020}, $33$: \cite{Atri-etal2020}, $34$: \cite{Trushkin-etal2018}, $35$: \cite{Khargharia-etal2010}, $36$: \cite{Miller-Jones-etal2009}, $37$: \cite{Plotkin-etal2019}, $38$: \cite{Corral-Santana-etal2016}, $39a$: \cite{Mata-Sanchez-etal2015}, $39b$: \cite{Shahbaz-etal2013}, $40$: \cite{Paice-etal2019}, $41$: \cite{Baby-etal2021}, $42$: \cite{Tetarenko-etal2021}, $43$: \cite{Russell-etal2019c}, $44$: \cite{Shafee-etal2006}, $45a$: \cite{Zhao-etal2021}, $45b$: \cite{Kushwaha-etal2021}, $46$: \cite{Stuchlk-Kolos2016}, $47$: \cite{Sreehari-etal2020}, $48$: \cite{Miller-etal2009}, $49$: \cite{Ludlam-etal2015}, $50$: \cite{Steiner-etal2013}, $51$: \cite{Wang-etal2018}, $52a$: \cite{King-etal2014}, $52b$: \cite{Pahari-etal2018a}, $53$: \cite{Miller-etal2018}, $54$: \cite{Guan-etal2021}, $55$: \cite{Walton-etal2017}
	\\
	\noindent$^\dagger$: Mass estimate of these sources are uncertain, till date. $^{\boxdot}$: VLA observation; $^{\boxtimes}$: ATCA observation; $^{\boxplus}$: VLBA observation in 2014.\\
	\noindent{\bf Note:} References for black hole mass ($M_{\rm BH}$), distance ($D$),  $F_\nu$ or $L_{\rm R}$, and spin ($a_{\rm k}$) are given in column 8 in sequential order. Data are complied based on the recent findings (see also \cite{Merloni-etal2003,Gultekin-etal2019}).
	\label{tab:GBH_Data}
\end{table*}

In the next section, we use equation (10) to estimate the shock luminosity ($L_{\rm shock}$) (equivalent to the kinetic power released by the disc) for black hole sources that include both BH-XRBs and AGNs. While doing this, the jets/outflows are considered to be compact as well as core dominated surrounding the central black holes. Further, we compare $L_{\rm shock}$ with the observed core radio luminosity ($L_{\rm R}$) of both BH-XRBs and AGNs.

\section{Astrophysical implications}

\label{sec:implication}

In this work, we focus on the core radio emission at $\sim 5$ GHz from the black hole sources in all mass scales starting from BH-XRBs to AGNs. We compile the mass, distance, and core radio emission data of the large number of sample sources from the literature.

\subsection{Source Selection: BH-XRBs}

We consider $16$ BH-XRBs whose mass and distance are well constrained, and the radio observations of these sources in LHS are readily available (see Table \ref{tab:GBH_Data}). The accretion in LHS \cite[]{Belloni-etal2005,Nandi-etal2012} is generally coupled with the core radio emission \cite[]{Fender-etal2004} from the sources. Because of this, we include the observation of compact radio emission at $\sim 5$ GHz to calculate the radio luminosity while excluding the transient radio emissions ({\it i.e.}, relativistic jets) commonly observed in soft-intermediate state (SIMS) \cite[see][and references therein]{Fender-etal2004,Fender-etal2009,Radhika-Nandi2014,Radhika-etal2016}. It may be noted that the core radio luminosity of some of these sources are observed at different frequency bands (such as $15$ GHz). For Cyg X-1, $15$ GHz radio luminosity was converted to $5$ GHz radio luminosity assuming a flat spectrum \cite[]{Fender-etal2000}, whereas for XTE J1118+480, we convert the $15$ GHz radio luminosity to $5$ GHz radio luminosity using a radio spectral index of $\alpha = +0.5$ considering $F_{\nu} = \nu^{\alpha}$ \cite[]{Fender-etal2001}. For these sources, we calculate $5$ GHz radio luminosity using the relation $L_{\rm R} \equiv \nu L_\nu= 4 \pi \nu F_5 D^2$ \cite[see][]{Gultekin-etal2019}, where $\nu \sim 5$ GHz, $F_5$ are the $\sim 5$ GHz flux, and $D$ is the distance of the source, respectively. It may be noted that our BH-XRB source samples differ from \cite{Merloni-etal2003} and \cite{Gultekin-etal2019} because of the fact that we use most recent and refined estimates of mass and distance of the sources under consideration, and accordingly we calculate their radio luminosity. Further, we exclude the source LS 5039 from Table \ref{tab:GBH_Data} as it is recently identified as NS-Plusar source \cite[]{Yoneda-etal2020}. In Table \ref{tab:GBH_Data}, we summarize the details of the selected sources, where columns {$1-8$ represent source name, mass, distance, spin, observation frequency ($\nu$), radio flux ($F_5$), core radio luminosity ($L_R$) and relevant references, respectively.

 \subsection{Source Selection: SMBH in AGN}
 
 We consider a group of AGN sources following \cite{Gultekin-etal2019} (hereafter G19) that includes both Seyferts and LLAGNs. For these sources, \cite{Gultekin-etal2019} carried out the image analysis to extract the core radio flux ($F_{\nu}$) that eventually renders their core radio luminosity ($L_R$). Here, we adopt a source selection criteria as (a) $M_{\rm BH}>  10^{5} M_\odot$ and (b) source observations at radio frequency $\nu \sim 5$ GHz, that all together yields $61$ source samples. Subsequently, we calculate the core radio luminosity of these sources as $L_R=4 \pi \nu F_{5}D^2$, where $F_5$ denote the core radio luminosity at $\nu = 5$ GHz frequency and obtain $L_R = 10^{32.5}-10^{40.8}$ erg s$^{-1}$. 
 
 
 Next, we use the catalog of \cite{Rakshit-etal2020} (hereafter R20) to include Supermassive black holes (SMBHs) in our sample sources. The R20 catalog contains spectral properties of $\sim 500,000$ quasars up to redshift factor $(z) \sim 5$ covering a wide range of black hole masses $10^7 - 10^{10} M_{\odot}$. The mass of the SMBHs in the catalog is obtained by employing the Virial relation where the size of the broad line region can be estimated from the AGN luminosity and the velocity of the cloud can be calculated using the width of the emission line. Accordingly, the corresponding relation for the estimation of SMBH mass is given by \cite[]{Kaspi-etal2000},
 $$
 \log \left( \frac{M_{\mathrm{BH}}}{M_{\odot}} \right) = a + b\log \left(\frac{\lambda L_{\lambda}}{10^{44} \mathrm{erg\,s^{-1}}} \right)  + 2\log \left( \frac{\mathrm{\Delta V}}{\mathrm{km\, s^{-1}}} \right),
 \eqno(11)
 $$
where $L_{\lambda}$ is the monochromatic continuum luminosity at wavelength $\lambda$ and $\Delta V$ is the FWHM of the emission line. The coefficients $a$ and $b$ are empirically calibrated based on the size-luminosity relation either from the reverberation mapping observations \cite[]{Kaspi-etal2000} or internally calibrated based on the different emission lines \cite[]{Vestergaard-Peterson2006}. Depending on the redshift, various combinations of emission line (H$\beta$, Mg II, C IV) and continuum luminosity ($L_{5100}$, $L_{3000}$, $L_{1350}$) are used. A detailed description of the mass measurement method is described in R20.   

The majority of AGN in R20 sample have $M_{\rm BH} > 10^8M_\odot$. As the low-luminosity AGNs (LLAGNs) with mass $M_{\mathrm{BH}}<10^7 M_{\odot}$ are not included in R20 sample, we explore the low-luminosity AGN catalog of \cite{Liu-etal2019} (hereafter L19). It may be noted that in L19, the black hole mass is estimated by taking the average of the two masses obtained independently from the H$\alpha$ and H$\beta$ lines.

In order to find the radio-counterpart and to estimate the associated radio luminosity, we cross-match both catalogs ($i.e.$, L19 and R20) with $1.4$ GHz FIRST survey \cite[]{White-etal1997} within a search radius of $2$ arc sec. The radio-detection fraction is $3.4 \%$ for R20 and $11.7\%$ for L19 AGN samples. We note that the present analysis deals with core radio emissions of black hole sources and many AGNs show powerful relativistic jets which could be launched due to Blandford-Znajek (BZ) process \cite[]{Blandford-Znajek1977} instead of accretion flow. Meanwhile,  \cite{Rusinek-etal2020} reported that the jet production efficiency of radio loud AGNs (RL-AGNs) is 10\% of the accretion disc radiative efficiency, while this is only 0.02\% in the case of radio quiet AGNs (RQ-AGNs) suggesting that the collimated, relativistic jets ought to be produced by the BZ mechanism rather than the accretion flow. Subsequently, we calculate the radio-loudness parameter ($R$, defined by the ratio of FIRST 1.4 GHz to optical $g$-band flux) and restrict our source samples for radio-quiet \cite[$R<19$; see][]{Komossa-etal2006} AGNs. As some radio sources are present in both catalogs ($i. e.$, L19 and R20), we exclude common sources from R20. With this, we find $1207$ and $911$ radio-quiet AGNs in the R20 and L19 AGN sample, respectively. Accordingly, the final sample contains $2118$ AGNs with black hole mass in the range 	$10^{5.1} < (M_{\rm BH}/M_\odot) < 10^{10.3}$.
 
The FIRST catalog provides $1.4$ GHz integrated radio flux ($F_{1.4}$), which is further converted to the luminosity $L_{1.4}$ (in watt/Hz) at $1.4$ GHz using the following equation as,
$$
L_{1.4} =  4\pi \times 10^{-7} \times \frac{D_L^2}{(1+z)^{(1+\alpha)}} \times F_{1.4},
\eqno(12a)
$$  
where we set the spectral index $\alpha=-0.8$ considering $F_\nu = \nu^{\alpha}$ \cite[]{Condon-1992} and $D_L$ refers the luminosity distance. Thereafter, we obtain the core radio luminosity 
$L_R$ at $5$ GHz adopting the relation \cite[]{Yuan-etal2018} given by,
$$
\log L_{R} = (20.9 \pm 2.1) + (0.77 \pm 0.08) \log L_{1.4}.
\eqno(12b)
$$
where $L_{R}$ is expressed in ${\rm erg~s}^{-1}$. The radio luminosity at $5$ GHz of our AGN sample has a range of $L_{R} = 10^{36.2}-10^{41.2}$ erg s$^{-1}$.

Following \cite{Rusinek-etal2020}, we further calculate the mean jet production efficiency of our sample and it is found to be only $\sim 0.02\%$ compared to the disc radiative efficiency. Such a low jet production efficiency suggests that the production of the jets in our sample is possibly due to accretion flow rather than the BZ process. Moreover, we calculate the $0.2-12$ keV X-ray luminosity ($L_x$) from the XMM-Newton data \cite[][3XMM-DR7]{Rosen-etal2016} for $119$ AGNs having both X-ray and radio flux measurements. The $L_x$ ranges from $1 \times 10^{41} - 2\times10^{46}$ erg s$^{-1}$ with a median of $10^{44}$ erg s$^{-1}$ . The ratio of X-ray ($0.2-12$ KeV) luminosity to radio luminosity ($L_R$ at $1.4$ GHz) has a range of $L_x/L_R \sim 1.5\times 10^2 - 6.6 \times 10^5$ with a median of $2.6\times 10^4$.
 
\subsection{Comparison of $L_{\rm shock}$ with Observed Core Radio Emission ($L_R$) of BH-XRBs and AGNs}
 
 \begin{figure*}
 	\vskip -4 cm
 	\begin{center}
 		\includegraphics[scale=0.9]{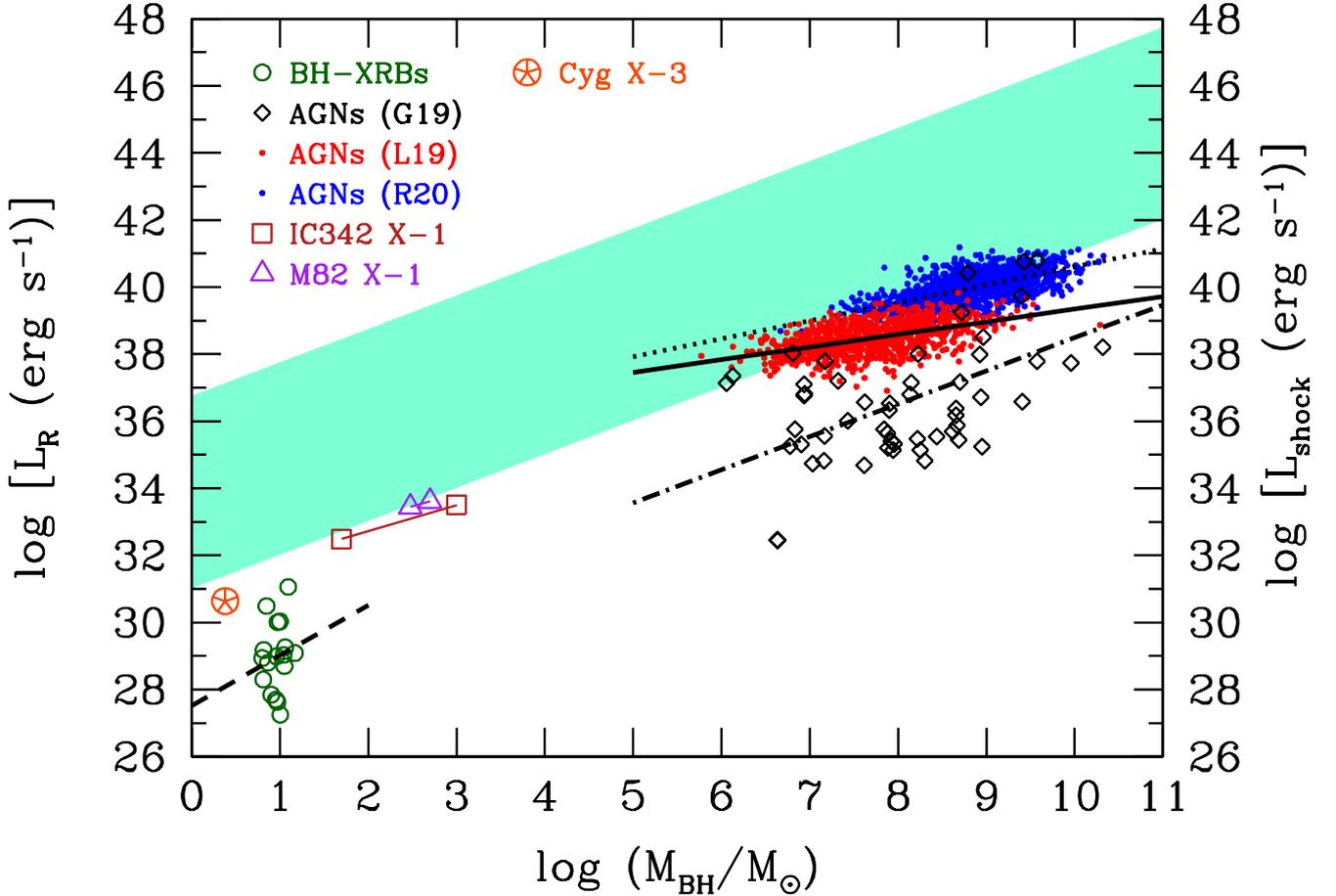}
 	\end{center}
 	\vskip -2 cm
 	\caption{Plot of kinetic power $L_{\rm shock}$ (in erg s$^{-1}$) released through the upper and lower surface of the disc due to the energy dissipation at the accretion shock as function of the central black hole mass ($M_{\rm BH}$). The same is compared with the observed core radio emission ($L_R$) of BH-XRBs and AGNs source samples. Shaded region (light-green) represents the model estimate of $L_{\rm shock}$ obtained for accretion rates $10^{-5}\lesssim {\dot m} \lesssim 1$ and $0 \le a_{\rm k} < 1$. Open circles denote BH-XRBs, whereas open diamonds, red dots and blue dots represent the AGN samples taken from \citet{Gultekin-etal2019}, \citet{Liu-etal2019} and \citet{Rakshit-etal2020}, respectively. Open squares and open triangles illustrate $L_R$ for IMBH sources. Solid, dotted, dot-dashed and dashed lines indicate the results obtained from liner regression for AGNs (L19), AGNs (R20), AGNs (G19), and BH-XRBs, respectively. See text for details.
 		}
 	\label{fig:M_LR_M03G19}
 \end{figure*}

 
 In Fig. {\ref{fig:M_LR_M03G19}}, we compare the shock luminosity (equivalently loss of kinetic power) obtained due to the energy dissipation at the shock with the observed core radio luminosities of central black hole sources of masses in the range $\sim 3-10^{10}M_\odot$. The chosen source samples contain several BH-XRBs and a large number of AGNs. In the figure, the black hole mass (in units of $M_\odot$) is varied along the $x$-axis, observed core radio luminosity ($L_R$) is varied along $y$-axis (left side) and shock luminosity ($L_{\rm shock}$) is varied along the $y$-axis (right side), respectively. We use $\Delta{\cal E}^{\rm max}$ calculated for black holes having spin range $0 \le a_{\rm k} \le 0.99$ (see Fig. \ref{fig:ak_dEmax}), to compute the shock luminosity $L_{\rm shock}$ which is analogous to the core radio luminosity ($L_R$) of the central black hole sources. Here, the radio core is assumed to remain confined around the disk equatorial plane ($\theta \sim \pi/2$) in the region $ r \le r_{\rm s}$. We vary the accretion rate in the range $10^{-5} \le {\dot m} \le 1$ to include both gas-pressure and radiation pressure dominated disc \cite[and references therein]{Kadowaki-etal2015} and obtain the kinetic power $L_{\rm shock}$ that is depicted using light-green color shade in Fig. {\ref{fig:M_LR_M03G19}}. The open green circles correspond to the core radio emission from the $16$ BH-XRBs while the dots and diamonds represent the same for AGNs. The black diamonds represent $61$ AGN source samples adopted from \cite{Gultekin-etal2019}. The red dots ($908$ samples) denote the low-luminosity AGNs (LLAGNs) \cite[]{Liu-etal2019} and the blue dots ($1207$ samples) represent the quasars \cite[]{Rakshit-etal2020}. At the inset, these three sets of AGN source samples are marked as AGNs (G19), AGNs (L19) and AGNs (R20), respectively. It is to be noted that we exclude Cyg X-3 from this analysis due to the uncertainty of its mass estimate and in the figure, we mark this source using red asterisk inside open circle. We carry out the linear regression analysis for (a) BH-XRBs, (b) AGNs (G19), (c) AGNs (L19), and (d) AGNs (R20) and estimate the correlation between the mass ($M_{\rm BH}$) and the core radio luminosity ($L_R$) of the black hole sources. We find that $L_{R} \sim M_{\rm BH}^{1.5}$ for BH-XRBs (dashed line), $L_{R} \sim M_{\rm BH}^{0.98}$ for AGNs (G19) (dot-dashed line), $L_{R} \sim M_{\rm BH}^{0.38}$ for AGNs (L19) (solid line), and $L_{R} \sim M_{\rm BH}^{0.54}$ for AGNs (R20) (dotted line), respectively. Fig. \ref{fig:M_LR_M03G19} clearly indicates that the kinetic power released because of the energy dissipation at the shock seems to be capable of explaining the core radio emission from the central black holes. In particular, the results obtained from the present formalism suggest that for ${\dot m} \lesssim 1$, only a fraction of the released kinetic power at the shock perhaps viable to cater the energy budget required to account the core radio emission for supermassive black holes although $L_R$ for stellar mass black holes coarsely follows shock luminosity ($L_{\rm shock})$.

It is noteworthy to mention that the radio luminosity of AGNs from G19 are in general lower compared to the same for sources from R20 and L19 catalogs. In reality, AGNs from L19 and R20 are mostly distant unresolved sources where it remains  challenging to separate the core radio flux from the lobe regions. Hence, a fraction of the lobe contribution is likely to be present in the estimation of their $L_R$ values even for radio quiet AGNs. Nonetheless, we infer that the inclusion of the L19 and R20 sources will not alter the present findings of our analysis at least qualitatively.

\subsection{$L_R$ for Intermediate Mass Black Holes}

 \begin{table*}
 	\caption{Physical and observational parameters of IMBH sources. }
 	\centering
 	\begin{tabular}{@{}lccccc} \hline \hline
 		Source Name  & Distance ($D$) & X-ray luminosity ($L_X$) & Mass Range ($M_{\rm BH}$)& Predicted Radio Luminosity($L_R$)  & References\\
 		&  (in Mpc) & (in erg s$^{-1}$) &   (in $M_\odot$)     &   (in erg s$^{-1}$)    & \\ \hline
 		IC342 X-1 & $3.93$ & $5.34\times 10^{39}$ & $50 - 1000$~$^\dagger$ & $3.09\times 10^{32} - 3.20 \times 10^{33}$& $1,2, 3$\\ \hline
 		M82 X-1 & $3.90$ & $2.00 \times 10^{40}$ & $300-500$ & $2.77 \times 10^{33} - 4.12 \times 10^{33}$ & $4, 5, 6$\\ \hline

 	\end{tabular}
 	\justify
 	\noindent{\bf References:} $1$: \cite{Tikhonov-Galazutdinova2010}, $2$: \cite{Agrawal-Nandi2015}, $3$: \cite{Cseh-etal2012}, $4$: \cite{Sakai-Madore1999}, $5$: \cite{Feng-Kaaret2009}, $6$: \cite{Pasham-etal2014}\\
 	\noindent$\dagger$: Mass estimate of this sources is uncertain, till date.\\
 	\noindent{\bf Note:} References for source distance ($D$), X-ray luminosity ($L_X$) and mass are given in column 6 in sequential order. 
 	
 	\label{tab:IMBH_Data}
 \end{table*}

The recent discovery by the LIGO collaboration resolves the long pending uncertainty of the possible existence of the intermediate mass black holes (IMBHs) \cite[]{Abbott-etal2020}. They reported the detection of IMBH of mass $142 ~M_\odot$ which is formed through the merger of two smaller mass black holes. This remarkable discovery establishes the missing link between the stellar mass black holes ($ M_{\rm BH} \lesssim 20 M_\odot $) and the Supermassive black holes ($ M_{\rm BH} \gtrsim 10^{6} M_\odot $). Due to limited radio observations of the IMBH sources, model comparison with observation becomes unfeasible. Knowing this constrain, however, there remains a scope to predict the radio flux for these sources by knowing the disc X-ray luminosity ($L_X$), source distance (D), and possible range of the source mass ($ M_{\rm BH} $). Following \cite{Merloni-etal2003}, we obtain the radio flux ($F_5$) at $5$ GHz using the relation given by,

\begin{align*}
F_{5} =& 10 \times \left( \frac{L_X}{3\times 10^{31}{\rm ~erg~s^{-1}}}\right)^{0.6} \times \left(\frac{M_{\rm BH}}{100 M_\odot}\right)^{0.78}\\
\times & \left(\frac{D}{10~{\rm kpc}}\right)^{-2}~\mu{\rm Jy}. \hskip 4.5 cm (13)
\end{align*}
Thereafter, using equation (13), we calculate $L_{\rm R} = 4\pi \nu F_5 D^2$ (see Table \ref{tab:IMBH_Data}). As a case study, we choose two IMBH sources whose $L_X$ and $D$ are known from the literature and examine the variation of $L_{\rm R}$ in terms of the source mass ($M_{\rm BH}$). Since the mass of IC 342 X-1 source possibly lie in the range of $50 \lesssim M_{\rm BH}/M_\odot \lesssim 10^3$ \cite[]{Cseh-etal2012,Agrawal-Nandi2015}, we obtain the corresponding $L_{\rm R}$ values which is depicted by the open squares joined with straight line in Fig. \ref{fig:M_LR_M03G19}. Similarly, we estimate $L_{\rm R}$ for M82 X-1 source by varying $M_{\rm BH}$ in the range $\sim 250-500 ~M_\odot$ \cite[]{Pasham-etal2014} and the results are presented by open triangles joined with straight line in Fig. \ref{fig:M_LR_M03G19}. Needless to mention that the predicted $L_{\rm R}$ for these sources reside below the model estimates. With this, we argue that the present model formalism is perhaps adequate to explain the energetics of the core radio emissions of IMBH sources.

\section{Discussion and Conclusions}

\label{sec:con}

In this paper, we study the relativistic, inviscid, advective, accretion flow around the black holes and address the implication of the dissipative accretion shock in explaining the core radio emissions from the  central engines. We observe that with the appropriate choice of the set of flow parameters, namely energy ($\cal E$) and angular momentum ($\lambda$), the global transonic accretion solutions pass through the shock discontinuity ($r_s$) around the black holes. When the shocks are considered to be dissipative ($i.e.$, accretion energy is being dissipated across the shock front) in nature, it reduces the local temperature that eventually decreases the post-shock pressure causing the shock front to settle down at smaller radii (see Fig. \ref{fig:r_M_dE}). Hence, the size of the PSC ($\sim r_s$) decreases as the level of energy dissipation ($\Delta \cal E$) is increased. We further point out that the shock induced global accretion solutions are generic solutions and such solutions are possible for wide range of ${\cal E}$ and $\lambda$ around weakly as well as rapidly rotating black holes (see Fig. \ref{fig:amda_Ein_Eout}). Subsequently, we calculate the maximum amount of accreting energy that can be extracted at the shock and find that ${\Delta \cal E} \sim 1\%$ for Schwarszchild black hole ($a_{\rm k}\rightarrow 0$) and ${\Delta \cal E} \sim 4.4\%$ for Kerr black hole ($a_{\rm k}\rightarrow 1$) (see Fig. \ref{fig:ak_dEmax}).

We implement our model formalism to explain the observed core radio emissions emanated from the vicinity of the black holes, in particularly when the compact core is yet to be separated from the central region. While doing this, we explore the entire range of the black hole masses starting from stellar mass to Suppermassive scale. We find that for $10^{-5} \lesssim {\dot m} \lesssim 1$, the dissipative shock model formalism is capable to account the energy budget associated with the core radio luminosity of large number of the central black hole sources particularly with $2176$ AGNs although $16$ BH-XRBs are also seen to comply sparsely. It appears that the present model estimate suffers overestimation from the radio luminosity of BH-XRB sources that perhaps causes the reticence of adopted model formalism. We also emphasize that one would get the degenerate $L_{\rm shock}$ due to the suitable combination of $\dot M$ and $\Delta {\cal E}$ as delineated in equation (10), which remains in broad agreement with $L_R$ (see Fig. \ref{fig:M_LR_M03G19}). In reality, $L_{\rm shock}$ is corroborated the core radio emissions from the region which is still not decoupled from the accretion disk to form jets and hence, $L_{\rm shock}> L_R$ seems to be not unrealistic as only a part of $L_{\rm shock}$ contributes to radio emission (other parts will be exhausted for (a) thermodynamical expansion and (b) against gravity). We further attempt to fill the missing link between the BH-XRBs and AGNs including two IMBH sources and predict $L_{\rm R}$ values as function of source mass ($M_{\rm BH}$) as their mass uncertainty is yet to be settled. We find that $L_{\rm R}$ (as function of $M_{\rm BH})$ for these IMBH sources resides inside the domain of the model estimates (see Fig. \ref{fig:M_LR_M03G19}) and therefore, we indicate that the plausible explanation of the core radio emission of these IMBH sources could be understood from this model formalism.

It is noteworthy to refer that there exists alternative scenarios involving magnetic fields where the rotational energy of the black hole is imparted to power the launching jets \cite[]{Blandford-Znajek1977}. On the contrary, a recent study indicates that the jet driving mechanism in all astrophysical objects possibly uses energy directly from the accretion disc, rather than black hole spin \cite[]{King-Pringle2021}. Moreover, it is inferred that jets from BH-XRBs are linked with the accretion states \cite[and references therein]{Fender-etal2009,Radhika-etal2016} indicating the launching of jets possibly happens from the accretion disc itself.

In addition, the accretion disk geometry is generally depends on ${\dot m}$ when radiative cooling processes are active inside the disk. However, in this work, we focus in examining the non-dissipative accretion flow and being transonic, flow must satisfy the regularity conditions. Because of these extra conditions, out of the three constants of the motions, namely, ${\cal E}$, $\lambda$, and ${\dot m}$, only two are sufficient ($i.e.$, ${\cal E}$ and $\lambda$) to obtain the accretion solutions \cite[]{Das-etal2001a} and therefore, the half-thickness ($H$) of the disk remains independent on $\dot m$.

We further indicate that the shock induced global accretion solutions are potentially promising in explaining the spectral state transitions of BH-XRBs \cite[]{Nandi-etal2018,Radhika-etal2018,Aneesha-etal2019,Baby-etal2020,Aneesha-Mandal2020}, when the two-component accretion flow configuration is espoused \cite[]{Chakrabarti-Titarchuk1995,Smith-etal2001,Smith-etal2002,Mandal-Chakrabarti2007,Iyer-etal2015}. With this, we further infer that the shock transition radius perhaps be visualized as the inner edge of the truncated disc \cite[]{Esin-etal1997,Done-etal2007,Kylafis-etal2012}.

Finally, we point out the limitations of the present model formalism. In our analysis, we ignore viscosity, magnetic fields and various radiative processes to avoid complexity, although these physical processes are likely to be relevant in the context of accretion physics. In addition, the accreting matter may dissipate both angular momentum and accretion rate from the post-shock region when jets/outflows are present \cite[]{Fukumura-Kazanas2007,Takahashi-Takahashi2010}. However, the mechanisms responsible for the mass loss and angular momentum loss from the disc still remain inconclusive. In addition, the estimation of mass loss generally depends on the outflow geometry which is again largely model dependent \cite[and references therein]{Chakrabarti1999,Das-etal2001b,Aktar-etal2015}. Hence, in this work, we only focus in studying the accretion flow for simplicity and therefore, the obtained shocked-induced global accretion solutions remain independent on accretion rate ($\dot m$) as radiative cooling processes are neglected, although $\dot m$ eventually regulates the estimation of $L_{\rm shock}$. Because of this, it remains theoretically infeasible to describe the physical states of accretion flow as well as its geometrical morphology around black holes while explaining the core radio luminosity by means of energy dissipation across the shock front. Considering all these issues, we argue that the overall findings of the present paper are expected to remain unaltered at least qualitatively. In addition, we point out that we restrict ourselves while carrying out the $L_R - M_{\rm BH}$ correlation analysis without invoking disc X-ray luminosity ($L_X$) due to the lack of X-ray observations for large number of chosen source samples. We plan to take up these issues for our future works and will be reported elsewhere.

\section*{Data availability statement}

The data underlying this article are available in the published literature.

\section*{Acknowledgments}
We thank the anonymous reviewers for constructive comments and useful suggestions that help to improve the quality of the paper. SD thanks Science and Engineering Research Board (SERB), India for support under grant MTR/2020/000331. SD also thank Department of Physics, IIT Guwahati for providing the infrastructural support to carry out this work. AN and SS thanks GH, SAG; DD, PDMSA and Director, URSC for encouragement and continuous support to carry out this research. IKD thanks the financial support from Max Planck partner group award at Indian Institute Technology of Indore (MPG-01).


\bsp

\label{lastpage}

\end{document}